\documentclass[useAMS,usenatbib]{mn2e}
\usepackage{epsfig}

\newcommand{\he}[1] {He\,{\sc #1}}
\newcommand{\hel}[2] {He\,{\sc #1}~$\lambda$#2}

\title[DW\,Cancri]{DW Cancri: a magnetic VY Scl star with an orbital
  period of 86 minutes.} 
\author[P. Rodr\'\i guez-Gil, B. T. G\"ansicke, S. Araujo-Betancor and
J. Casares]{P. Rodr\'\i guez-Gil$^{1}$\thanks{E-mail:
prguez@astro.soton.ac.uk}, B. T. G\"ansicke$^{1}$,
S. Araujo-Betancor$^{1}$ and J. Casares$^{2}$\\ $^{1}$School of
Physics and Astronomy, University of Southampton, Southampton SO17
1BJ, UK\\ $^{2}$Instituto de Astrof\'\i sica de Canarias, V\'\i a
L\'actea, s/n, La Laguna, E-38200, Santa Cruz de Tenerife, Spain\\ }
\begin{document}

\date{Accepted 2003. Received 2003; in original form 2003}

\pagerange{\pageref{firstpage}--\pageref{lastpage}} \pubyear{2003}

\maketitle

\label{firstpage}

\begin{abstract}
We present the first time-resolved spectroscopic study of the
cataclysmic variable DW\,Cancri. We have determined an orbital period
of $86.10\pm0.05$\,min, which places the system very close to
the observed minimum period of hydrogen-rich cataclysmic variables. This
invalidates previous speculations of DW Cnc being either a permanent
superhumper below the period minimum or a nova-like variable with an
orbital period longer than 3 hours showing quasi-periodic
oscillations. The Balmer and \he{i} lines have double-peaked profiles
and exhibit an intense S-wave component moving with the orbital
period. Remarkably, the Balmer and \he{i} radial velocity curves are
modulated at two periods: $86.10\pm0.05$\,min (orbital) and $38.58\pm0.02$\,min. The same short period is found in the equivalent width variations of the
single-peaked \hel{ii}{4686} line. We also present time-resolved
photometry of the system which shows a highly-coherent variation at
38.51 min, consistent with the short spectroscopic period. The large
number of similarities with the short-period intermediate polar
V1025\,Cen lead us to suggest that DW\,Cnc is another intermediate
polar below the period gap, and we tentatively identify the
photometric and spectroscopic 38\,min signals with the white dwarf
spin period. DW\,Cnc has never been observed to undergo an outburst,
but it occasionally exhibits low states $\sim 2$ mag fainter than its
typical brightness level of $V\simeq14.5$, resembling the behaviour of
the high mass-transfer VY\,Scl stars.

\end{abstract}


\section{Introduction}
Despite their name, a surprisingly large fraction of cataclysmic
variables (CVs, see \cite{warner95-1} for a review) do not display
frequent and/or large amplitude brightness variations and may easily
escape detection. DW\,Cnc (FBS\,0756+164) is an example of such
systems. It has been identified in the First Byurakan Survey, a
spectroscopic galaxy survey, as a CV because of its noticeable Balmer
emission lines. \cite{stepanian82-1} reported brightness variations in
the range $V\approx 15-17.5$ and suggested that DW\,Cnc is a dwarf
nova, a classification later supported by \cite{kopylovetal88-1}.

\cite{uemuraetal02-2} performed the first time-resolved photometric
study of DW\,Cnc and found it at $R=14.7$. They also observed a variation
with an amplitude of $\sim 0.3$ mag\, ---which they explained as quasi-periodic oscillations
(QPOs)---\, and no outburst activity. Two possible periods (37.5 and 73.4 min) were reported. The authors attributed the variability to either permanent superhumps with
the longest period or trapped disc oscillations in a 3-hour period
nova-like system.\par

We recently noticed DW\,Cnc as a bright but poorly studied CV in the
course of our ongoing large-scale spectroscopic survey for new CVs
(e.g. \citealt{gaensickeetal00-2}; G\"ansicke, Hagen \& Engels
\citeyear{gaensickeetal02-2}; \citealt{araujo-betancoretal03-2}).  In
this paper we present the first time-resolved spectroscopic
observations of DW\,Cnc, along with the analysis of new and archival
photometry.

\section{Observations and data reduction}
\subsection{Photometry}
Differential CCD photometry of DW\,Cnc was obtained with the 1.0-m
Jacobus Kapteyn Telescope on La Palma (Isaac Newton Group archive
data; 1999 February 24, 28) and the 0.82-m IAC80 telescope on Tenerife
(2003 April 12). The JKT observations were performed in the $V$-band
using the $1024 \times 1024$ pixel$^2$ Tek4 CCD detector, whilst
unfiltered images were taken with the $1024 \times 1024$ pixel$^2$
Thomson CCD at the IAC80 (See Table~\ref{table1}).  The raw images
were debiased and flat-fielded in the standard way. After aligning all
images, we measured the instrumental magnitudes of the target and two
close comparison stars using the point spread function (PSF) packages
within {\sc iraf}\footnote{{\sc iraf} is distributed by the National
  Optical Astronomy Observatories.}. Differential photometry of
DW\,Cnc was then computed relative to each comparison star. The
accuracy of the photometric measurements was $\la 0.07$ mag and $\sim
0.1$ mag for the JKT and IAC80 data, respectively. A magnitude offset
between the JKT $V$-band data and the IAC80 white-light data was
calculated from the instrumental magnitudes of the same comparison
stars measured in both image sets. The IAC80 light curve was then
shifted by this offset in order to be consistent with the $V$-band
data. The fact that the JKT and IAC80 light curves share approximately
the same mean level indicates that the colour correction is negligible.

\begin{center}
\begin{table}
 \begin{minipage}{70mm}
  \caption{\label{table1} Log of photometric observations}
  \begin{tabular}{@{}lcccc@{}}
  \hline
  Date                      & Telescope & Band        & Exp. (s)    & Coverage (h)\\
  \hline
  1999 Feb 24               & JKT &    $V$ & 120 &  5.58\\
  1999 Feb 28               & JKT &    $V$ & 90 &  3.03\\
  2003 Apr 12               & IAC80 &    White light & 15 &  1.52\\
\hline
  \end{tabular}
\end{minipage}
\end{table}
\end{center}

\subsection{Spectroscopy}
Time-resolved spectroscopy of DW\,Cnc was performed with the Calar
Alto 2.2-m telescope (2003 April 7, 8, 9, 11) and the 2.5-m Isaac
Newton Telescope (INT) on La Palma (2003 March 13 and April 22,
26). At the 2.2-m telescope we used the Calar Alto Faint Object
Spectrograph ({\sc cafos}) along with the G-100 grism. This setup,
together with a slit width of 1.2\arcsec, gave a spectral resolution
of 5.7 \AA~(FWHM) and a useful wavelength range of
$\lambda\lambda4255-8300$ on the $2048 \times 2048$ pixel$^2$ SITe
CCD. The Intermediate Dispersion Spectrograph ({\sc ids}) on the INT,
with the R632V grating and a slit width of 1.5\arcsec, provided a
useful range of $\lambda\lambda4400-7150$ at 2.3 \AA~(FWHM) resolution
on the $2048 \times 4100$ pixel$^2$ EEV10a chip (April data). For the
March run at the INT we used the R900V grating and a slit width of
1.2\arcsec. This setup gave a spectral range of
$\lambda\lambda3800-5500$ at a resolution of 1.6 \AA~(FWHM). Details
on the spectroscopic observations are found in Table~\ref{table2}.\par

\begin{figure}
\begin{center}
  \mbox{\epsfig{file=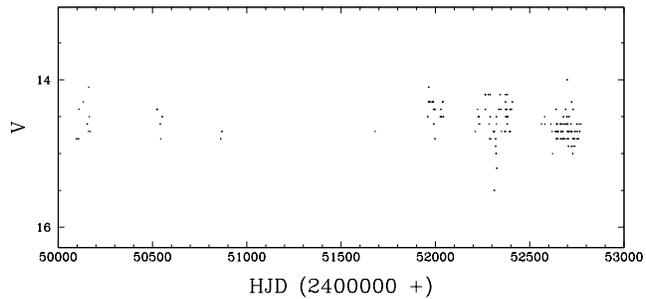,angle=-90,width=8.5cm}}
\end{center}
\caption{\label{fig1b} Long-term, $V$-band light curve of DW\,Cnc
from the AFOEV archive and the VSNET. The measurements span
approximately 8 years. No outburst has been recorded.}
\end{figure}

The $V$-band magnitude of DW\,Cnc was measured during the Calar Alto
run from the \textsc{cafos} acquisition images with respect to the
Hubble Space Telescope guide star GSC0136301286 ($V=13.82$). The values are
$V=13.98\mathrm{,}~14.14\mathrm{,}~14.35\mathrm{,~and~}14.25$, for the
four nights. The brightness of DW\,Cnc is therefore consistent with
the mean magnitude measured over 8 years (see Fig.~\ref{fig1b}).\par

The individual spectra were corrected for the bias level and
flat-field structure. After sky-subtraction, they were optimally
extracted \citep{horne86-1}. For the wavelength calibration a
low-order polynomial was fitted to the arc data, the {\sl rms}
being always less than a tenth of the spectral dispersion. The
wavelength scale for each target spectrum was interpolated from the
wavelength scales of two neighbouring arc spectra. These reduction
processes were performed using the standard packages for long-slit
spectra within {\sc iraf}.\par

Only the {\sc cafos} spectra ---which have the widest wavelength
coverage--- were flux calibrated. The spectrophotometric standard
BD\,+75$^{\circ}$ 325 observed with the same setup in order to correct
for the instrumental response. The flux calibration, as well as all
subsequent analysis on the whole data set were performed using Tom
Marsh's {\sc molly} package.

\begin{center}
\begin{table}
 \begin{minipage}{70mm}
  \caption{\label{table2} Log of spectroscopic observations}
  \begin{tabular}{@{}lcccc@{}}
  \hline
  Date                      & Telescope     & Exp. (s)    & Coverage (h)\\
  \hline
  2003 Mar 12               & INT   & $6 \times 600$ &  1.05\\
  2003 Apr 07               & 2.2-m & $6 \times 600$ &  1.00\\
  2003 Apr 08               & 2.2-m & $6 \times 600$ &  0.99\\
  2003 Apr 09               & 2.2-m & $3 \times 600$ &  0.37\\
  2003 Apr 11               & 2.2-m & $5 \times 600$ &  0.95\\
  2003 Apr 22               & INT   & $22 \times 300$ &  2.07\\
  2003 Apr 26               & INT   & $6 \times 300$, $14 \times 180$ &  1.36\\
\hline
  \end{tabular}
\end{minipage}
\end{table}
\end{center}

\section{Photometric variability}
\subsection{\label{s-longterm}Long-term behaviour}
Fig.~\ref{fig1b} shows the long-term light curve of DW\,Cnc. The
photometric measurements were taken from the AFOEV
archive\footnote{ftp://cdsarc.u-strasbg.fr/pub/afoev/} and the
VSNET\footnote{http://www.kusastro.kyoto-u.ac.jp/vsnet}.  A mean
magnitude of $V \sim 14.3$ is derived from the long-term light curve,
with excursions of up to $\sim \pm 0.5$ mag. This scatter is
comparable with the short-term variation of the source (see
below). No outburst has been recorded in the $\sim 8$-year monitoring
of DW\,Cnc. However, the system occasionally displays faint states
down to at least $V \sim 16$. This is confirmed by the USNO-A2.0
measurements, which caught DW\,Cnc at $R=16.5$ and $B=16.6$, whereas
the DSS1 and DSS2 plates show DW\,Cnc at its typical brightness of $V
\sim 14-15$. The magnitude range covered by the long-term light curve
in Fig.\,\ref{fig1b} is consistent with that given by
\cite{stepanian82-1}. Taken at face value, the long-term behaviour of
DW Cnc resembles that of the VY\,Scl stars, which have high mass
accretion rates and are found above the period gap.

\begin{figure}
\begin{center}
  \mbox{\epsfig{file=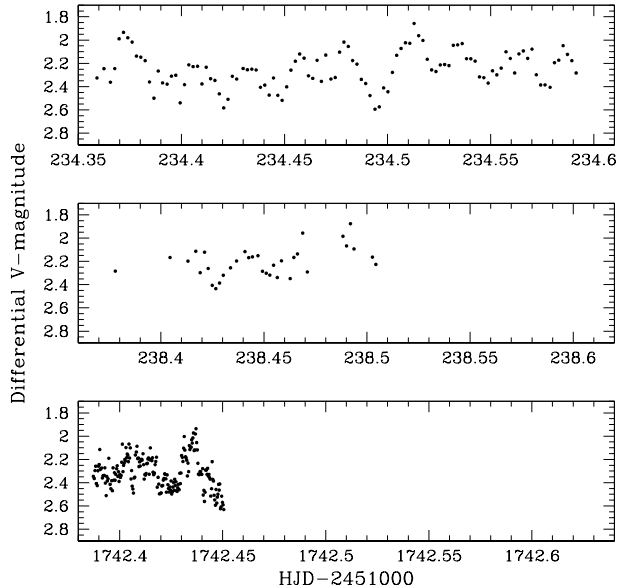,width=8.5cm}}
\end{center}
\caption{\label{fig1} $V$-band ({\em top} and {\em middle} panels) and
unfiltered ({\em bottom}) light curves of DW\,Cnc. A magnitude shift
has been applied to the latter according to the same comparison stars
in $V$-band and white light. See text and Table~\ref{table1} for
details.}
\end{figure}

\subsection{The JKT and IAC80 light curves}  
The JKT and IAC80 light curves of DW\,Cnc clearly show short
time-scale variability with a peak-to-peak amplitude of $\approx
0.3-0.4$ mag (Fig.\,~\ref{fig1}). Inspection of the light curves by
eye suggests a separation between consecutive maxima of $\sim 40$ min.

\begin{figure}
\begin{center}
  \mbox{\epsfig{file=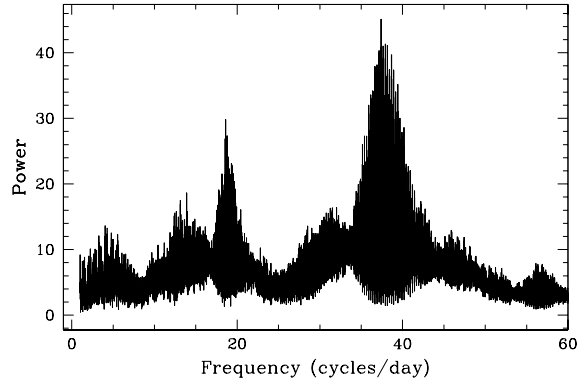,width=8cm}}
\end{center}
\caption{\label{fig4} AOV periodogram of the photometric data. Two
prominent peaks can be seen at approximately 38.5 and 77.4 min.}
\end{figure}

To explore the presence of periodicities in the data we computed an
analysis-of-variance \citep[AOV,][]{schwarzenberg-czerny89-1}
periodogram of the whole photometric data set, which is shown in
Fig.~\ref{fig4}. The strongest peak is centred at $\approx 38$
min. The very complex alias structure of the periodogram (due
to the coarse sampling) prevents an unequivocal determination of
the maximum peak frequency to be made. $\chi^2$ sine fits to the entire data set folded over the alias signals contained in the range 35.37 to 39.38
cycles/day favour 37.39 cycles/day ($= 38.51 \pm 0.02$ min; the
error being half the FWHM of the peak). The periodograms also show significant power at around $77.37 \pm 0.03$ min. The ratio between the two
strongest frequencies detected in the photometry is approximately 2
but, again, the inadequate sampling does not allow to address a definite
conclusion as to whether the two frequencies are comensurate or not.
\cite{uemuraetal02-2} found two peaks in their power spectrum at 37.5
and 73.4\,min, but their periodograms also display strong aliasing
effects due to poor sampling. A more extensive photometric study
is then fundamental to test in detail whether both periods are
independent or not.\par

\begin{figure}
\begin{center}
  \mbox{\epsfig{file=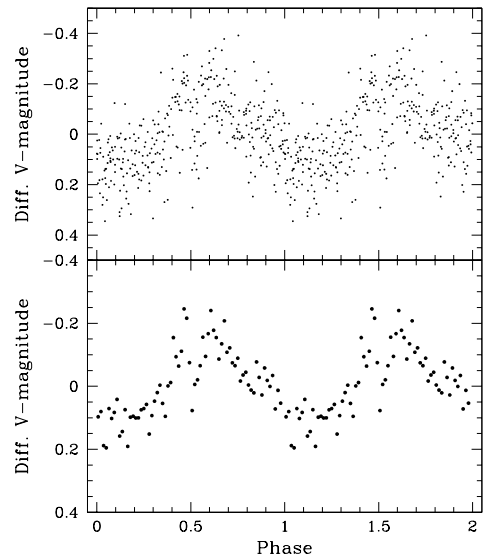,width=8cm}}
\end{center}
\caption{\label{fig5} Folded light curve on the 38.51-min period. In
the bottom panel the data have been averaged into 70 phase bins. The
curves are shown twice for clarity.}
\end{figure}

Fig.\,\ref{fig5} shows all the photometric data folded on 38.51 min
and averaged into 70 phase bins. The high degree of coherence in the
photometric signal is remarkable, since it is detected at the same
frequency (within the errors) in data spanning more than 4 years. This
strongly suggests that its origin is closely related to a ``stable
clock'' in the system, such as the orbital period or the spin period
of the white dwarf.

Noticeably, the light curve exhibits a dip of $\sim 0.4~\mathrm{mag}$
at approximately maximum brightness, appearing every 38.51
minutes. The dips can be easily seen in the IAC80 light curve (bottom
panel of Fig.~\ref{fig1}). The total duration of one of these events
is $\sim 5$ min, which makes it unlikely to be an eclipse by the
secondary star. If that was the case the orbital period of DW Cnc
should be 38.51 min, very much shorter that the observed minimum orbital period
of CVs with non-degenerate secondaries and inconsistent with our
spectroscopic study described below.

\section{The average spectrum and line profile variations} 
\label{s-average_spectrum}
\begin{figure}
\begin{center}
  \mbox{\epsfig{file=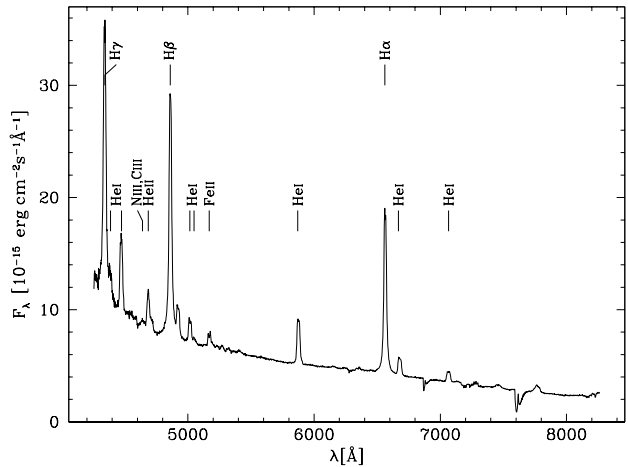,angle=-90,width=8.5cm}}
\end{center}
\caption{\label{fig6} Average spectrum of DW\,Cnc from the CAFOS data (Calar Alto Observatory).}
\end{figure}

In Fig.~\ref{fig6} we show the average of the 20 \textsc{cafos}
spectra. The spectrum of DW\,Cnc is dominated by Balmer and \he{i}
emission lines superimposed on a blue continuum. There is noticeable
\hel{ii}{4686} emission, which indicates the presence of a source of
ionising photons in the system. The Bowen blend of C/N and several
lines of Fe\,{\sc ii} are also present. There are no obvious
absorption features from either the secondary or the white dwarf. In
Table~\ref{table3} we show some line parameters as measured in the
average spectrum.

We have extended the spectral energy distribution (SED) of
DW\,Cnc into the infrared using its 2MASS magnitudes ($J=14.65 \pm
0.03$, $H=14.34 \pm 0.05$, $K=14.03 \pm 0.06$). A fit to the combined
optical/IR SED of DW\,Cnc (after converting the IR magnitudes into
fluxes) gives $F_\lambda \propto \lambda^{-2.4}$, which is exactly the
spectral slope of a steady-state, optically thick accretion disc
\citep{lynden-bell69-1}.

\begin{figure}
\begin{center}
  \mbox{\epsfig{file=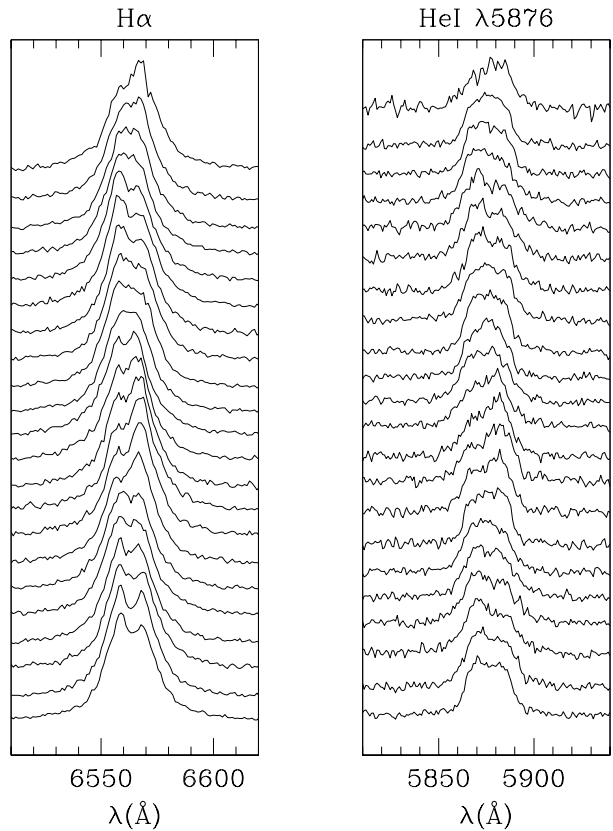,width=8.8cm}}
\end{center}
\caption{\label{fig7} Evolution of the H$\alpha$ and \hel{i}{5876}
profiles observed on April 22. Time runs from bottom to top. Note the changing relative intensity of
the peaks.}
\end{figure}

\begin{figure*}
\begin{center}
  \mbox{\epsfig{file=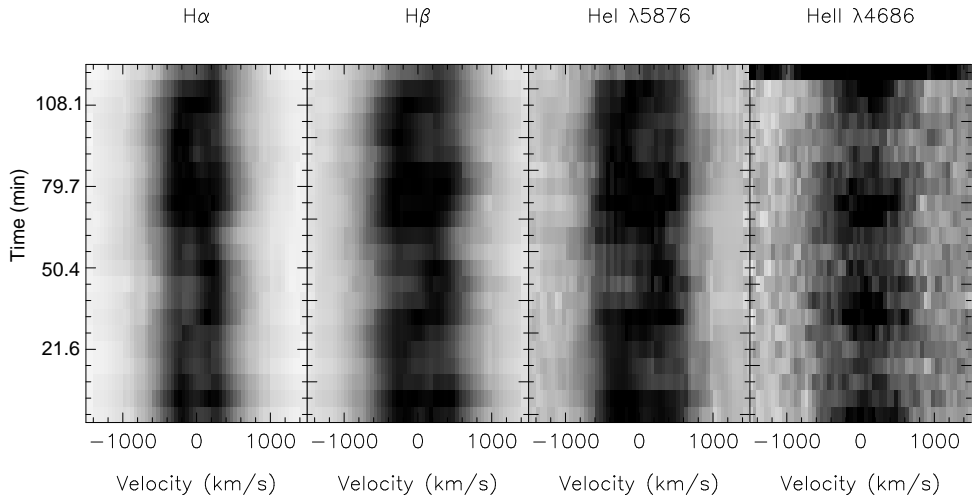,width=13cm}}
\end{center}
\caption{\label{fig8} H$\alpha$, H$\beta$, \hel{i}{5876} and
\hel{ii}{4686} trailed spectra diagrams of the April 22 data. An
emission S-wave can be seen moving inside the double-peaked profiles
with the exception of \hel{ii}{4686}. Note the ``flaring'' of
\hel{ii}{4686} on a time scale of $\simeq40$\,min. Black means
emission.}
\end{figure*}

The line profiles, with the exception of \hel{ii}{4686}, are
clearly double-peaked ($\sim 10$ \AA~separation in H$\alpha$),
indicating the presence of an accretion disc. Fig.~\ref{fig7} presents the evolution of the H$\alpha$ and
\hel{i}{5876} profiles for the night of 2003 April 22. The relative
intensity of the two peaks clearly changes with time. At some phases
the lines turn single-peaked, showing a nearly flat-topped
profile. Furthermore, the profiles display asymmetric wings extending
up to $\sim 2000$ km s$^{-1}$. All of this suggests the presence of at
least another component in the Balmer and \he{i} lines such as an
emission S-wave moving accross the double-peaked profile, as well as
emission from material moving very close to the primary. To
investigate this in more detail we constructed trailed spectrograms of
the April 22 H$\alpha$, H$\beta$, \hel{i}{5876} and \hel{ii}{4686}
lines after normalizing the spectra with a spline fit to the
continuum (see Fig.\,\ref{fig8}). An emission S-wave with a
semi-amplitude of $\sim 200-300$ km s$^{-1}$ is indeed detected with
the same phasing in H$\alpha$, H$\beta$ and \hel{i}{5876}. This value is consistent with the velocity
semi-separation of the double peak ($\sim 230$ km s$^{-1}$ in
H$\alpha$), suggesting that the emission S-wave originates in the
outer regions of the disc, possibly at the location of the bright
spot.

\begin{table}
 \centering
 \begin{minipage}{70mm}
  \caption{\label{table3} Line parameters measured in the average spectrum}
  \begin{tabular}{@{}lccc@{}}
  \hline
  Line                   & Centre        & EW    & FWHM          \\
                            & (km~s$^{-1}$) & (\AA)     & (km~s$^{-1}$) \\
  \hline
  H$\alpha$                  & $34\pm8 $    & $96\pm1$ & $1046\pm18$  \\
  H$\beta$                  & $30\pm11 $    & $79\pm1$ & $1530\pm17$  \\
  H$\gamma^\dagger$\footnote[0]{$^\dagger~~$From 2003 March data.}                 & $17\pm5$    & $64\pm1$ & $1659\pm7$  \\ 
  H$\delta^{\dagger}$                 & $118\pm7$   & $48\pm1$ & $1662\pm10$  \\
  He\,{\sc ii} $\lambda$4686~~~~~~~~ & $66\pm91$    & $9\pm1$ & $1472\pm163$   \\
He\,{\sc i} $\lambda$5876  & $44\pm32$  &  $25\pm1$ & $1313\pm46$  \\
 He\,{\sc i} $\lambda$4472  & $106\pm46$    &  $17\pm1$ & $1444\pm87$  \\
  \hline
  \end{tabular}
\end{minipage}
\end{table}

Remarkably, the strength of the Balmer, \he{i} and \hel{ii}{4686}
lines suffers a significant increase around spectra number 14 (see
Fig.~\ref{fig8}), which is probably caused by enhanced emission since
the continuum is not seen to significantly drop in the non-normalized spectra. Similar enhancements seem to take place around spectra number $1-2$ and 7. These ``flares'' are more
evident in the \hel{ii}{4686} trailed spectra, which does not
show the S-wave observed in the Balmer and \he{i} lines. The flares
repeat every $\sim 40$ min, a time scale comparable with the
photometric variability. The same behaviour is observed in the spectra
taken on April 26.

\section{The orbital period of DW\,Cnc}

\subsection{Radial velocity variations\label{rvc}}
Visual inspection of our trailed spectrograms (Fig.\,\ref{fig8})
reveals an S-wave modulated at a a period of $\sim
0.05-0.06~\mathrm{d}=72-86.4~\mathrm{min}$.  If we identify, by
analogy to most other CVs, the observed velocity variation of the
S-wave with the orbital revolution of the system, the orbital
period of DW\,Cnc is very close to the observed period minimum for hydrogen-rich
CVs.\par

\begin{figure*}
\begin{center}
  \mbox{\epsfig{file=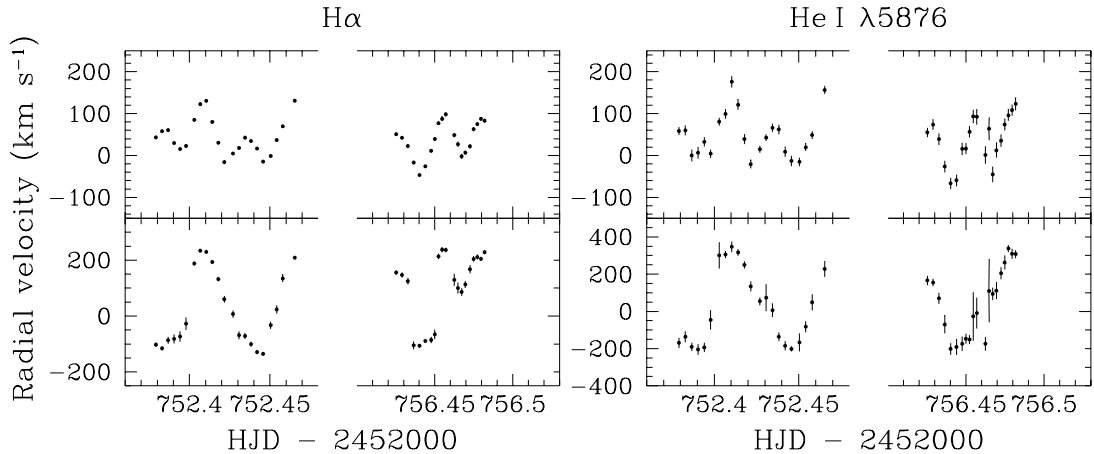,width=6cm,angle=-90}}
\end{center}
\caption{\label{fig10} H$\alpha$ (left panel) and \hel{i}{5876} (right panel) radial velocity curves from the April 22 and April 26 INT spectra. The curves at the top of each panel were obtained by correlating the line profiles with a Gaussian having the FWHM of the corresponding line in the average spectrum. For the bottom curves a Gaussian template of $\mathrm{FWHM}=300$ km s$^{-1}$ was employed.}
\end{figure*}

We have measured the radial velocities of H$\alpha$ and \hel{i}{5876}
for all the \textsc{cafos} and \textsc{ids} spectra. We first rebinned
the normalized spectra to a constant velocity scale for each line. The
radial velocities were then derived by correlating the individual
profiles with a single Gaussian template. The FWHM of the Gaussian
function was chosen either as the FWHM of the corresponding line
measured in the average spectrum (see Table~\ref{table3}), or fixed to
$\mathrm{FWHM}=300$ km s$^{-1}$. We will refer to the two FWHM in the
following as ``broad'' and ``narrow'', respectively.  The correlation
with the narrow Gaussian is intended to follow the motion of the
S-wave, which is much narrower than the double-peaked line
profile. The resulting radial velocity curves are shown in
Fig.\,\ref{fig10}.\par

The radial velocities obtained by correlating with the broad Gaussian
display an unusual morphology, as maxima and minima do not have the
same velocity. In the curves corresponding to the April 22 data high-
and low-velocity maxima seem to alternate.  The separation between
consecutive maxima is $\sim 0.03$\,d for all the lines, very close to
the main photometric period. A different picture arises when the
radial velocities are measured by correlation with the narrow
Gaussian. Now, the radial velocity curves seem to trace the motion of the S-wave (especially in the April 22 data). They are modulated at a period of $\sim 0.05-0.06$ d, approximately twice the period obtained when
using the broad Gaussian template, i.e. within the range in which we expect the orbital period of DW\,Cnc to be.

\begin{figure}
\begin{center}
  \mbox{\epsfig{file=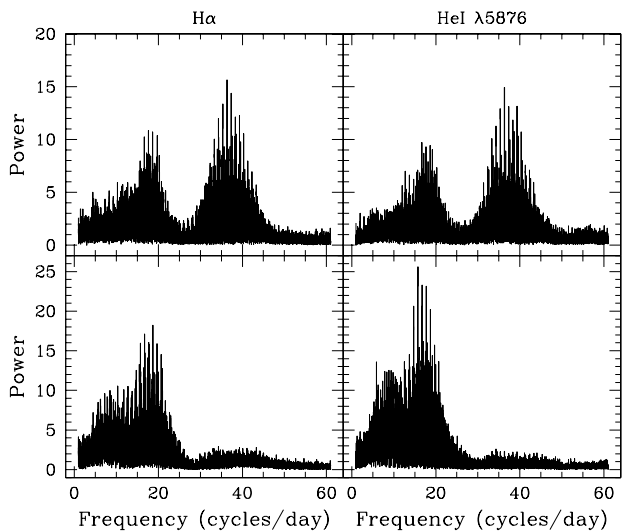,width=8.8cm}}
\end{center}
\caption{\label{fig13} AOV periodograms of the H$\alpha$ and
\hel{i}{5876} radial velocity curves. {\em Top panels}: correlation
with the broad Gaussian template. {\em Bottom panels}: correlation with the narrow Gaussian template having $\mathrm{FWHM}=300$ km s$^{-1}$.}
\end{figure}

We computed AOV periodograms for all the H$\alpha$ and \hel{i}{5876}
radial velocity curves. These are shown in Fig.~\ref{fig13}. The
periodograms reflect the behaviour observed in the double- and
single-humped radial velocity curves. The correlation with the broad
Gaussian results in periodograms with most of the power around 36
cycles/day, whilst those measured using the narrow Gaussian have the
strongest signal at about half this frequency. Unfortunately, the
complex alias structure of the periodograms prevents us to select
an unique value of the ``best period''. Table~\ref{table4} lists the
periods of the strongest peaks and their neighbour aliases in order of
decreasing power.

In view of the information extracted from the radial velocities we
strongly favour an orbital period in the range $0.054-0.060$\,d. The
unusual shape of the radial velocity curves obtained from the
correlation with the broad Gaussian seems to point to the
existence of at least another emission component which contaminates
the radial velocity curves. Hence, the possibility of having the
velocities modulated at two different periods can not be discarded.

\subsection{\label{s-vr} V/R ratios}
\begin{figure*}
\begin{center}
  \mbox{\epsfig{file=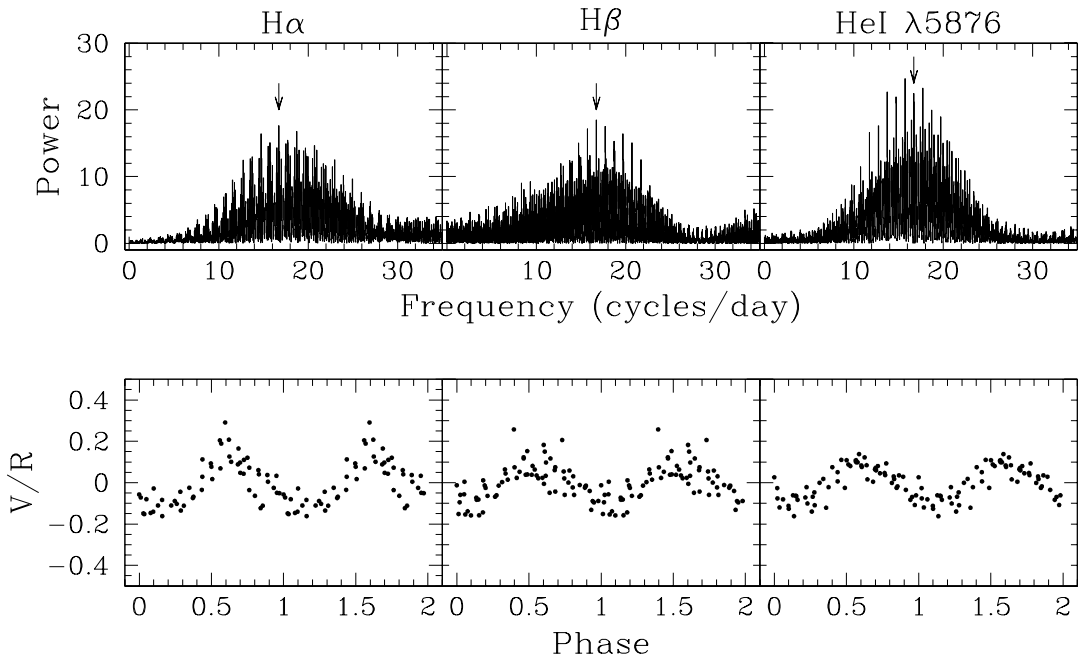,width=9cm,angle=-90}}
\end{center}
\caption{\label{fig_vr} {\em Top panel}: AOV periodograms of the
H$\alpha$, H$\beta$, and \hel{i}{5876} $V$/$R$ ratios. The arrows mark
the alias at 86.10 min, for which the smoothest folded curves are
found. {\em Bottom panel}: $V$/$R$ ratios folded on the 86.10-min
period. A whole cycle has been repeated for continuity.}
\end{figure*}

So far, the most reliable piece of information for the determination of the orbital period we have is the motion of the S-wave, which is especially prominent in \hel{i}{5876} and is very like to come from the bright spot. The effect of the S-wave on the double-peaked line profiles is such that we expect the ratio between the blue and red peak intensities to be modulated at the orbital period. We have computed these $V$/$R$ ratios for the H$\alpha$, H$\beta$, and \hel{i}{5876} profiles. The periodograms of the $V$/$R$ curves are displayed in Fig.~\ref{fig_vr}.

The cleanest modulation is obtained when folding $V$/$R$ curves on $86.10 \pm 0.05$ min. The quoted error
represents an average value of half the FWHM of the corresponding peak
in the periodograms. This period is consistent with one of the aliases
seen in the periodograms of the radial velocity curves obtained by
correlation with the narrow Gaussian template (see
Table~\ref{table4}). The H$\alpha$ velocities showed the best
smoothest modulation at 76.94 min, which is the
second strongest alias observed in the H$\alpha$ $V$/$R$
ratios. Based on our spectroscopic data, we favour 86.1\,min as
the orbital period of DW\,Cnc, even though we can not exclude the
$-1$\,d$^{-1}$ and $+1,+2$\,d$^{-1}$ aliases.

Shortly before the submission of this paper, we became aware of a more extensive spectroscopic study carried out by J. Thorstensen (private communication). He finds two {\em distinct} periods in his radial velocity measurements which are very close to our best estimate of the orbital period (based on the $V/R$ ratios) and to the strongest peak detected in our photometry.

\begin{table}
 \centering
 \begin{minipage}{70mm}
  \caption{\label{table4} Radial velocities periodogram analysis}
  \begin{tabular}{@{}llcc@{}}
  \hline
  Line                       & Method        & Period    & Aliases         \\
                             &  & (min)     & (min) \\
  \hline
  H$\alpha$                ~~  &  $\mathrm{FWHM}=1046$ km s$^{-1}$                       & 39.60 & 41.90  \\
                           ~~  &                                     &        & 38.59 \\
  H$\alpha$                ~~  &  $\mathrm{FWHM}=300$ km s$^{-1}$    & 76.94 & 86.12  \\
                           ~~  &                                     &        & 81.26  \\
  \hel{i}{5876}            ~~  &  $\mathrm{FWHM}=1313$ km s$^{-1}$                  & 39.60  & 40.75  \\
                           ~~  &                                     &        & 38.59 \\
  \hel{i}{5876}            ~~  &  $\mathrm{FWHM}=300$ km s$^{-1}$    & 91.58  &  86.12 \\
                           ~~  &                                     &        & 81.26 \\

  \hline
  \end{tabular}
\end{minipage}
\end{table}

\section{\hel{ii}{4686} equivalent widths}
\begin{figure}
\begin{center}
  \mbox{\epsfig{file=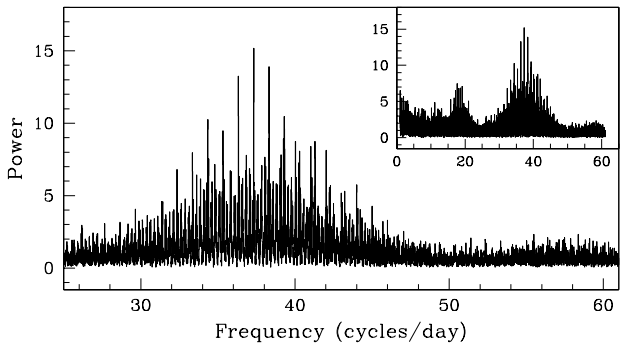,width=8.8cm}}
\end{center}
\caption{Periodogram of the \hel{ii}{4686} equivalent width curve. The strongest peak is centred at 38.58 min.}
\label{fig14}
\end{figure}

The \hel{ii}{4686} trailed spectra presented in Fig.~\ref{fig8} show
that this line experiences flares which repeat in a periodic
fashion. The separation between two consecutive flares is
consistent with the photometric period, so we decided to compute the
EW of the \hel{ii}{4686} and perform a period analysis. The EW was
measured in the line velocity interval $\left(-900,800\right)$ km
s$^{-1}$ to avoid contamination by the nearby Bowen blend and
\hel{i}{4713} line. The AOV periodogram of the resulting EW curve is
shown in Fig.~\ref{fig14}. The strongest peak is located at $38.58 \pm
0.02$ min, which is consistent with what is found in the photometry and
Thorstensen's radial velocities.\par

Our photometric and spectroscopic data then seem to reflect the
presence of two independent modulations. The fact that they are also
found in the radial velocities suggests that line emission is
produced in different locations. A modulation of the radial
velocities with the spin period of the white dwarf is observed in most
intermediate polar (IP) CVs which suggests that the 38.58-min oscillation might
represent (or might be related to) the spin period of the white dwarf.

\begin{figure}
\begin{center}
  \mbox{\epsfig{file=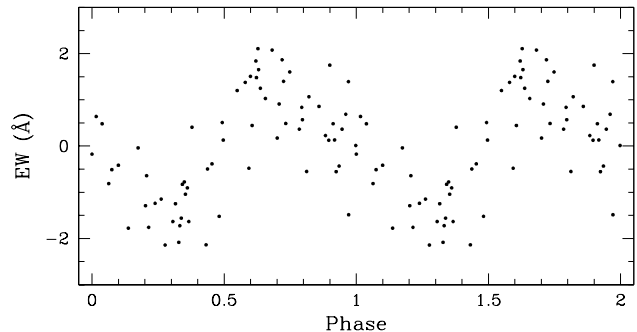,width=8.8cm}}
\end{center}
\caption{\hel{ii}{4686} equivalent width curves folded on the 38.58-min period. A full cycle has been repeated for continuity.}
\label{fig15}
\end{figure}

\section{Discussion}

\subsection{DW Cnc as a VY Scl star}
We have shown that DW\,Cnc has a likely orbital period of 86.10\,min,
a value very close to the observed period minimum of
hydrogen-rich CVs. This observed period minimum has been very
early explained by \citep{paczynski81-1} as the consequence of the
secondary star turning from a (nearly) main-sequence star into a
degenerate object and the corresponding change of its mass-radius relation.
However, models including more detailed stellar physics
notoriously predict the minimum period to be near 65\,min, which is in
strong disagreement with the observations. The mismatch between theory
and observations can be resolved if an additional angular momentum
loss mechanism besides gravitational is active in short period CVs
\citep[e.g.][]{patterson98-1,kingetal02-1}.

The orbital period range $\sim 70-120$\,min is
populated primarily by SU\,UMa dwarf novae and by polars (CVs
containing a strongly magnetic white dwarf which prevents the
formation of a disc). A fair fraction ($\sim$ 15 per cent) of the dwarf novae belongs to
the WZ\,Sge sub-class, showing infrequent but huge amplitude
outbursts. As far as we are aware, no outburst of DW\,Cnc has ever
been recorded (Section\,\ref{s-longterm},
Fig.\,\ref{fig1b}). Combining the ``normal'' brightness of DW\,Cnc
($V\sim 14-15$) with the large number of all-sky surveys and amateur
observers scanning the sky it appears rather unlikely that possible
outbursts of DW\,Cnc have been overlooked. On the contrary, there is
multiple evidence that DW\,Cnc displays occassional \textit{low
states}, dropping by $\sim 2$\,mag below its typical brightness. This
behaviour is reminiscent of the VY\,Scl stars, which are high
mass-transfer CVs found above the period gap, typically in the $3-4$~h
period range. Combining the long-term behaviour of DW\,Cnc with the
fact that the slope of its optical/IR spectral energy distribution is
consistent with that of a steady-state accretion disc
(Section\,\ref{s-average_spectrum}), we suggest that the mass transfer
rate in this system must be unusually high for its orbital period,
keeping its accretion disc persistently in the hot, ionised state.\par

Recently, \cite{hameury02-1} have suggested that the absence of
outbursts in the VY Scl stars during low state (when the mass transfer
rate is believed to be sufficiently low for thermal disc instabilities
to occur) could be due to the truncation of the accretion disc by the
primary's magnetic field. This hypothesis is compatible with the
results presented in this paper, as DW\,Cnc clearly shows VY\,Scl-like low
states and the coherent oscillations detected in its optical light
curve are very likely due to the rotation of a magnetic white
dwarf.\par

Other CVs commonly showing VY\,Scl-like low states are the SW Sex
stars (see e.g. \citealt{martinez-paisetal99-1};
\citealt{tayloretal99-1}; and Rodr\'\i guez-Gil \& Mart\'\i nez-Pais
\citeyear{rodriguez-gil+martinez-pais02-1} for an updated review of
the properties of the SW Sex stars). These CVs have been recently
proposed as magnetic accretors based on the detection of variable
circular polarization and emission-line flaring
(\citealt{rodriguez-giletal01-1}), which gives further support to the
idea of the VY Scl states being caused by a combination of high
mass-transfer rate and magnetic truncation of the inner disc.

\subsection{A possible intermediate polar scenario}
The optical spectrum of DW\,Cnc differs from those of other
short-period dwarf novae by the presence of an unusually strong
\hel{ii}{4686} line, indicating the presence of high-energy
photons in the system. Moreover, \hel{ii}{4686} exhibits a remarkable
behaviour. It does not display the S-wave seen in the Balmer and
\he{i} lines and its EW is modulated at a period of 38.58
min (see Fig.~\ref{fig14}). If the 86.10-min modulation represents the
actual orbital period (and there is strong evidence for that), the
\hel{ii}{4686} modulation is not likely to be related to the orbital
motion. The same modulation is seen in the photometry which we have
shown to be highly coherent. Hence, the 38.58-min period must be
related to another stable clock in the system, most likely to
the asynchronous rotation of a magnetic white dwarf.\par

It is interesting to note that the photometric and spectroscopic
behaviour of DW\,Cnc is similar in many aspects to that of V1025 Cen
\citep{buckleyetal98-1}, a confirmed intermediate polar CV. This system
has an orbital period of 84.6\,min and a spin period of 35.8\,min
(nearly half the orbital period). The optical spectrum of V1025\,Cen
is nearly identical to that of DW\,Cnc, containing strong
multi-component emission lines. The radial velocity curves of
V1025\,Cen reveal two periodicities, reflecting both the spin and the
orbital motion. Our identification of two different periodicities in the radial velocity curves of DW Cnc at 38.58 and 86.10 minutes points to the same origin. Unfortunately, our photometric and spectroscopic data are not sufficient to unequivocally pinpoint the complex interplay between orbital, spin and sideband periods.\par

An important difference with V1025\,Cen is that DW\,Cnc has not been
detected in the ROSAT All Sky Survey (Voges et~al. 2000\nocite{vogesetal00-1}). The
galactic column density of neutral hydrogen in the direction of
DW\,Cnc is not very high ($\simeq2.9\times10^{20}\mathrm{cm^{-2}}$),
making it unlikely that the non-detection is due to absorption. It
may, however, be that DW\,Cnc was in one of its low states during
the ROSAT observation, and therefore too faint to be detected by the satellite. A detailed study aiming at the detection of X-rays modulated at the suspected
white dwarf spin period is mandatory for the confirmation of DW\,Cnc
as an IP.

\section{Conclusions}

In what follows we summarize our results:

\newcounter{millo}
\begin{list}{\arabic{millo}.}{\usecounter{millo}}
\item The emission lines of DW Cnc have a double-peaked profile, indicating the
presence of a moderately-inclined accretion disc in the system. The
Balmer and \he{i} lines show an intense S-wave moving within the
double peaks. None of the stellar components is detected.
\item We have found the orbital period of DW\,Cnc to be most
likely $P_\mathrm{orb}=86.10 \pm 0.05$\,min, which places DW\,Cnc very
close to the observed minimum period of hydrogen-rich CVs.
\item The Balmer and \he{i} radial velocity curves show a double-periodic morphology with modulations both at the orbital period and 38.58 min. In addition, the EW of the \hel{ii}{4686} line is also modulated at the latter period .
\item The photometric light curves also show a coherent modulation at 38.51 min, consistent with the short period found in the radial velocities and the \hel{ii}{4686} EWs. We suggest that this variation is produced by the asynchronous rotation of a magnetic white dwarf.
\item No dwarf nova outburst has ever been observed in DW\,Cnc. Instead, the system displays low states during which its brightness
drop by $\sim 2$ mag, resembling the VY Scl stars. The optical/IR
spectral energy distribution of DW\,Cnc is very similar to that of a
steady-state accretion disc. Combining all evidences, DW\,Cnc appears
to be a high-mass transfer system below the period gap.
\item Despite the lack of significant X-ray emission from DW Cnc, the
observed behaviour (both photometric and spectroscopic) is very
similar to that of the IP V1025\,Cen, and we suggest that DW\,Cnc is another
short-period intermediate polar.
\end{list} 

Despite its brightness, DW Cnc remains a poorly studied CV. Although
we have provided evidence that points to a magnetic nature, it must be
confirmed by means of more extensive photometry and higher-resolution
spectroscopy. UV, X-ray, and polarimetric observations will also
provide fundamental information.

\section*{Acknowledgments}
We thank the anonymous referee for his/her valuable comments on the original manuscript. We are very grateful to John Thorstensen for sharing his results with us. PRG, BTG and SAB thank PPARC for support through a PDRA grant, an
Advanced Fellowship, and a postgraduate studentship, respectively.
The Isaac Newton Telescope and the Jacobus Kapteyn Telescope are operated on the island of La Palma by
the Isaac Newton Group in the Spanish Observatorio del Roque de los
Muchachos of the Instituto de Astrof\'\i sica de Canarias. The IAC80
telescope is operated on the island of Tenerife by the Instituto de
Astrof\'\i sica de Canarias in the Spanish Observatorio del Teide of
the Instituto de Astrof\'\i sica de Canarias. Based in part on data
obtained at the German-Spanish Astronomical Centre, Calar Alto,
operated by the Max-Planck-Institute for Astronomy, Heidelberg,
jointly with the Spanish National Commission for Astronomy.  We acknowledge the use of data from the Isaac Newton Group archive. The use
of the {\sc molly} and {\sc trailer} packages developed
by Tom Marsh is also acknowledged.

\bibliographystyle{mn2e}
\bibliography{aamnem99,$HOME/tex/aabib}

\bsp

\label{lastpage}

\end{document}